\documentclass[fleqn,10pt]{wlscirep}
\usepackage{authblk}
\usepackage[utf8]{inputenc}
\usepackage[T1]{fontenc}    
\usepackage{hyperref}       
\usepackage{url}            
\usepackage{booktabs}       
\usepackage{amsfonts} 
\usepackage{amsmath}
\usepackage{nicefrac}       
\usepackage{microtype}  
\usepackage{lipsum}		
\usepackage{graphicx}
\usepackage{comment}
\usepackage{doi}
\usepackage{xcolor}
\usepackage{caption}
\usepackage{subcaption}
\usepackage{parskip}

\begin{document}
\title{Topological, or Non-topological? A Deep Learning Based Prediction}

\author[1,*]{Ashiqur Rasul}
\author[2]{Md Shafayat Hossain}
\author[1]{Ankan Ghosh Dastider}
\author[1]{Himaddri Roy}
\author[2]{M. Zahid Hasan}
\author[1,+]{Quazi D. M. Khosru}
\affil[1]{Department of Electrical and Electronic Engineering, Bangladesh University of Engineering and Technology, Dhaka,1205, Bangladesh}
\affil[2]{Department of Physics, Princeton University, Princeton, NJ, 08544, USA}

\affil[*]{ashiqurrasul@ug.eee.buet.ac.bd}
\affil[+]{qdmkhosru@eee.buet.ac.bd}

\begin{abstract}
Prediction and discovery of new materials with desired properties are at the forefront of quantum science and technology research. A major bottleneck in this field is the computational resources and time complexity related to finding new materials from \emph{ab initio} calculations. In this work, an effective and robust deep learning-based model is proposed by incorporating persistent homology and graph neural network which offers an accuracy of $91.4\%$ and an F1 score of $88.5\%$ in classifying topological vs. non-topological materials, outperforming the other state-of-the-art classifier models. The incorporation of the graph neural network encodes the underlying relation between the atoms into the model based on their own crystalline structures and thus proved to be an effective method to represent and process non-euclidean data like molecules with a relatively shallow network. The persistent homology pipeline in the suggested neural network is capable of integrating the atom-specific topological information into the deep learning model, increasing robustness, and gain in performance. It is believed that the presented work will be an efficacious tool for predicting the topological class and therefore enable the high-throughput search for novel materials in this field. 
\end{abstract}

\flushbottom
\maketitle

\thispagestyle{empty}
\section{Introduction}
Recent years have witnessed the rise of machine learning, enabling a multitude of new applications, ranging from accelerated drug discovery to personalized advertising \cite{vamathevan2019,ekins2019exploiting, CHOI2020175}. Machine learning and more specifically deep learning-based techniques are becoming popular in material science, thanks to their high accuracy, computational speed, and ease of use in comparison with the \emph{ab initio} calculations \cite{schmidt2019recent}. Although our computational power has increased manifold, the task of discovering new materials, and exploring their types, properties, and structures is still significantly time-consuming and computationally expensive \cite{jha2019enhancing,chibani2020machine,jiang2021topological}. On the other hand, deep learning is capable of providing the same result several orders of magnitude faster than traditional ones \cite{peano2021rapid,giustino20212021}. The key requirement to apply deep learning-based methods is the availability of a large-scale data set. Thankfully, density functional theory \cite{kohn1965self} based computational databases comprising \emph{ab initio} and symmetrical calculations have been available in recent years, and it enables the application of deep learning in material science \cite{jha2019enhancing,tao2021machine,faber2017prediction,xue2016accelerated,ward2017including,legrain2017chemical,stanev2018machine,ramprasad2017machine,seko2017representation}.

Data-driven intelligent models learn certain features or descriptors of the materials from the provided data and can make certain decisions based on those in an automated system \cite{chibani2020machine,lecun2015deep,chen2019graph}. Encoding the properties, i.e., structural and other relevant information of various materials is a critical step for building an adroit deep learning model. In fact, the objective of how to represent the atoms of a molecule led to a surge of recent interest in graph convolutional neural networks \cite{wu2020comprehensive, fung2021benchmarking}. Notably, Xie et al. \cite{xie2018crystal} developed a generalized crystal graph convolutional neural network (CGCNN) to predict material properties by embedding molecular information into graph neural network and it led to further developments in this field \cite{chen2019graph, park2020developing, louis2020graph, karamad2020orbital}. For example, Ref. \cite{karamad2020orbital} incorporated atomic orbital interaction features in the developed graph neural network model, outperforming the CGCNN one. Although various machine learning and deep learning models for predicting material properties can be found in the literature, a sophisticated framework that can predict a particular class of material is scarce. Here we build such a model focusing on the topological class of materials. The discovery of topological classification of materials introduced us to unprecedented new physics and predicting new topological materials remains a major technological goal \cite{moore2010birth,qi2011topological}. There are a few frameworks geared towards this frontier: Claussen et al.'s \cite{claussen2020detection} model built upon a gradient-boosted trees algorithm \cite{friedman2001greedy} can predict the DFT-computed topology of a given material. Acosta et al.'s \cite{acosta2018analysis} compressed sensing-based statistical-learning approach can create a two-dimensional map where trivial insulators and Quantum spin-Hall insulators are separated in different domains. The deep neural network presented in Ref. \cite{sun2018deep} can predict the topological invariant of one-dimensional four-band and two-dimensional two-band insulators based on momentum space Hamiltonian. Hamiltonian is also used as an input in Ref. \cite{zhang2018machine}, where the supervised neural network is trained to distinguish various topological phases of topological band insulators.


Deep learning-based automatic prediction of the topological materials requires an efficient representation of the material and a scheme to extract the useful features for the network. Persistent homology, which explores the qualitative features based on the geometry and topology \cite{otter2017roadmap}, is a strong candidate for topological data analysis. It has already found numerous applications in scientific and engineering applications including biological information analysis \cite{cang2018integration,gameiro2015topological}, prediction of chemical stability \cite{xia2015persistent}, and crystalline compound representation \cite{jiang2021topological}. For modeling crystals, the inclusion of many-atom interaction is crucial. Atom-specific persistent homology (ASPH) can capture such interactions and thereby extract the true topological information of the materials, thus providing strong features for the deep learning model \cite{jiang2021topological}. In this work, we integrate ASPH with the graph convolutional neural network to build an efficient classification model for predicting the topological class of material. We find that the use of atom-based chemical information as the learning feature boosts the overall classification performance significantly. We first benchmark the performance of our model with well-known materials and then inspect a few recently discovered materials to justify our claim. Satisfactory performance in both 2-class (topological vs. nono-topological/trivial) and 3-class (topological insulator, semimetal, and trivial) classifications, and relevant comparison with other studies prove the effectiveness of the proposed model.

\section{Results}
\subsection{Model Architecture}
Our graph neural network – persistent homology ensemble model architecture encodes the structural, chemical, and topological information of the material. The graph representation of the crystals acts as a structural descriptor whereas the persistent homological feature vector acts as a topological descriptor. Therefore, the model can be divided into two separate but operating in parallel models: one network works with the graph representation learning and the other part performs the algebraic homological operation on the material crystal structures to produce a topological feature vector. These two representations in two different spaces are finally combined to produce the complete representation of the whole crystal system. This conjoined feature vector is then fed into a deep neural network to produce the final prediction result. The entire mechanism of merging homological and graph features to reach the final result is presented in Fig. 1.
\vspace{5pt}
\begin{figure}[h!]
\centering\includegraphics[width=12cm]{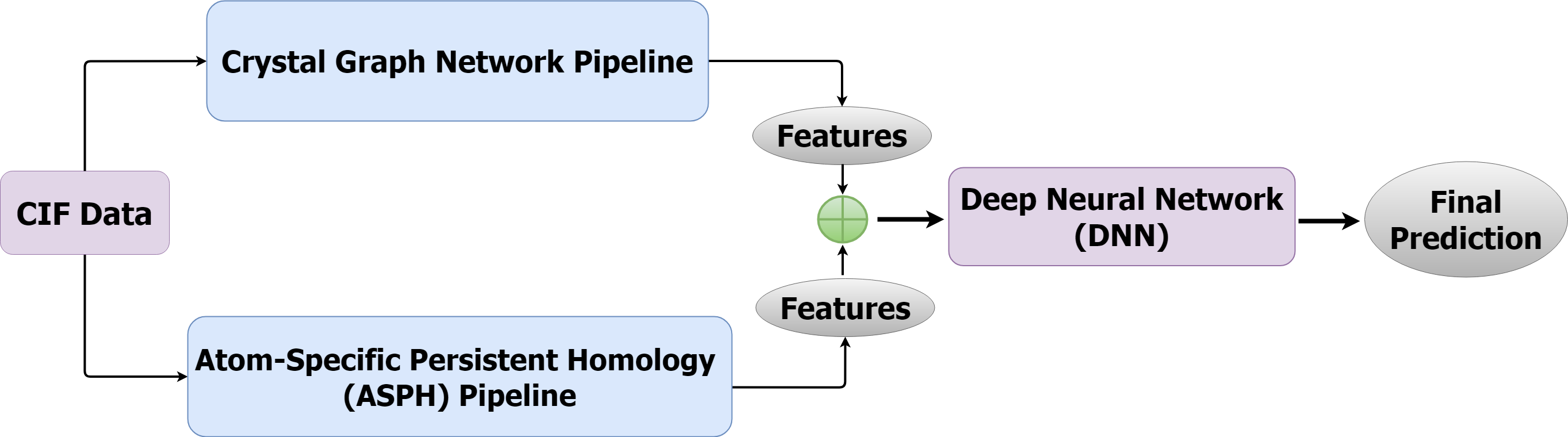}
\vspace{10pt}
\caption{Schematic of the composite network executed in our model. Features are extracted from the simultaneously progressing graph neural networks and atom-specific persistent homology pipelines. In turn, these features are concatenated to feed the predictor deep neural network model to ensure proper utilization of all the necessary features supplied by the two pipelines and thereby achieve the best possible prediction result from the proposed deep neural network architecture.}
\label{GCN-ASPH Composite Network}
\end{figure}

\subsubsection{Crystal Graph Generation}
The graph representation of data can bear more vigorous information on the basic structure thanks to its ability to represent the relation between the entities. The basic components of a graph are nodes and edges- here the edges create connections between the nodes. When considering a molecule as a graph, each atom is represented as a node in the graph while the interactions among the atoms are represented as edges. Thus atoms and their bonding information can be encoded into a graph structure. However, the graphs are non-structured and non-euclidean in nature, so accessing and working with the information is challenging and requires scaling down to a lower dimension. A key advantage of a graph convolutional neural network is that it can work with varying and non-ordered node connections. It is, therefore, significantly advantageous over the traditional convolutional neural network which only works with the regular structured Euclidean data.

\begin{figure}[h!]
\centering\includegraphics[width=15cm]{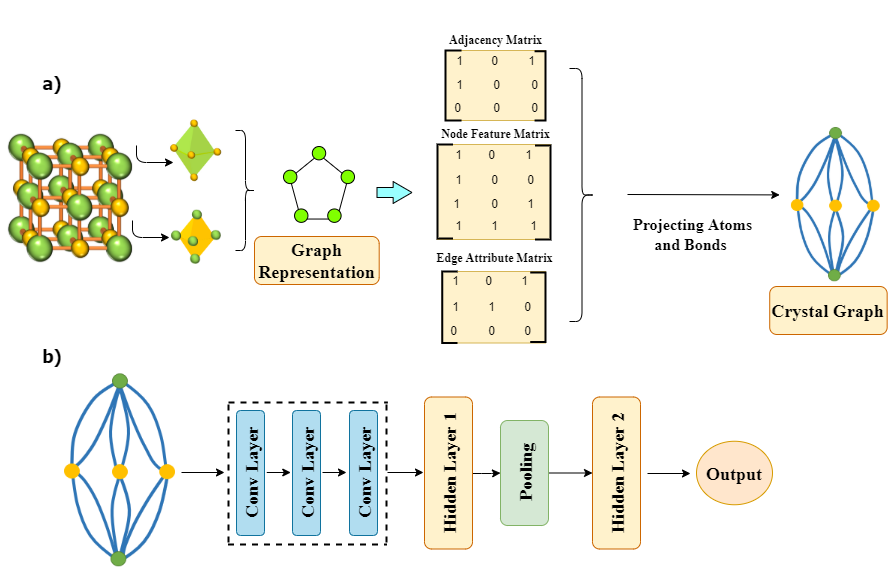}
\caption{Crystal graph convolutional neural network (CGCNN) pipeline used in this study. (a) A molecule is represented by a crystal graph, consisting of an adjacency matrix, edge, and node attribute matrix. Here the atoms of a crystal form the nodes, whereas the bonds are represented by edges. (b) Crystal graph is propagated through a series of graph convolutional layers followed by a graph pooling layer, and lastly, a fully connected layer to produce the desired output.}
\label{cgcnn}
\end{figure}

As illustrated in Fig. \ref{cgcnn}, we implement the graph neural network portion of our model to be a graph classification network that adopts a graph convolutional network algorithm. Its basic framework \cite{xie2018crystal} entails a transformation of the input atomic information into a graph where the nodes and edges represent the atoms and the bonds, respectively \cite{fung2021benchmarking}. Leveraging these graphs, a convolutional neural network is built that is capable of predicting the material topological class just like a classification problem.  The network contains convolutional and pooling layers where a convolutional layer applies filters on the input vector to capture the accurate position of features whereas the pooling layer creates a reduced (pooled) size layer with precise focus on the useful features. The pooling layer also makes the model tolerant to the little distortions at the input side. For the properties of $i^{th}$ atom (represented by node i), a feature vector $v_{i}$ is formed. Similarly, the edge is embedded with the $u(i,j)_{k}$ feature vector. Here $k^{th}$ bond connects the atoms i and j. At the training stage, the node feature vectors are updated as \cite{xie2018crystal}:

\begin{equation}
v_{i}^{(t+1)} = v_{i}^{(t)} + \sum_{j,k}^{} [\sigma(z_{{(i,j)}_{k}}^{(t)}\textbf{W}_{f}^{(t)})+\textbf{b}_{f}^{(t)})] \odot g(z_{{(i,j)}_{k}}^{(t)}\textbf{W}_{s}^{(t)})+\textbf{b}_{s}^{(t)}).
\end{equation}

Here $\odot$ and $\sigma$ denote element-wise multiplication and sigmoid function, respectively. $z_{{(i,j)}_{k}}^{(t)} = v_{i}^{(t)} \oplus v_{j}^{(t)} \oplus u(i,j)_{k}$ incorporates the neighbouring vectors through concatenation. $\textbf{W}^{t}$ is the weight vector and $\textbf{b}^{t}$ represents the bias on $t^{th}$ convolution stage. A detailed description of CGCNN can be found in Ref. \cite{fung2021benchmarking}.

Each node is initiated with a feature vector of dimension 64. In the subsequent three stages of the convolutional network, each nodal feature vector is updated through gradient propagation based on the aggregation of the neighboring atoms. The model allows a maximum of 12 neighbors for individual atoms at a maximum radius of 15 {\AA}. The convolutional layers consist of a series of linear (fully connected layers) with sigmoid activation function and batch normalization layers, followed by the pooling layers intended for dimensionality reduction. These features are then fed to a single-stage hidden layer of dimension 128, eventually producing output features extracted from material graph information.

\subsubsection{Atom-Specific Persistent Homology (ASPH)}
In order to build a functional neural network for predicting topological classes of materials, we need to extract the necessary  features related to the complex crystal geometry as well as the interactions between atoms. However, features provided by the traditional topological descriptors are not sufficient to depict the complete crystal structure because they are over-simplified and lack the required geometrical information to train a neural network. It necessitates a more accurate crystalline representation and feature extraction approach capable of portraying the overall crystal domain interactions. 

We find persistent homology to be an effective tool for this purpose. Persistent homology can successfully encode the multi-scale crystalline geometric information as well as the topological invariant. Besides crystal geometry, it is necessary to consider the atomic diversity and crystal periodicity information to distinguish one crystal from the others of a similar kind. Here we choose atom-specific persistent homology (ASPH) because it can provide element-level chemical information and adaptations to periodicity necessary to capture the discrepancies of crystal structures \cite{asph}.

\begin{figure}[h!]

\centering\includegraphics[width=1.0\linewidth]{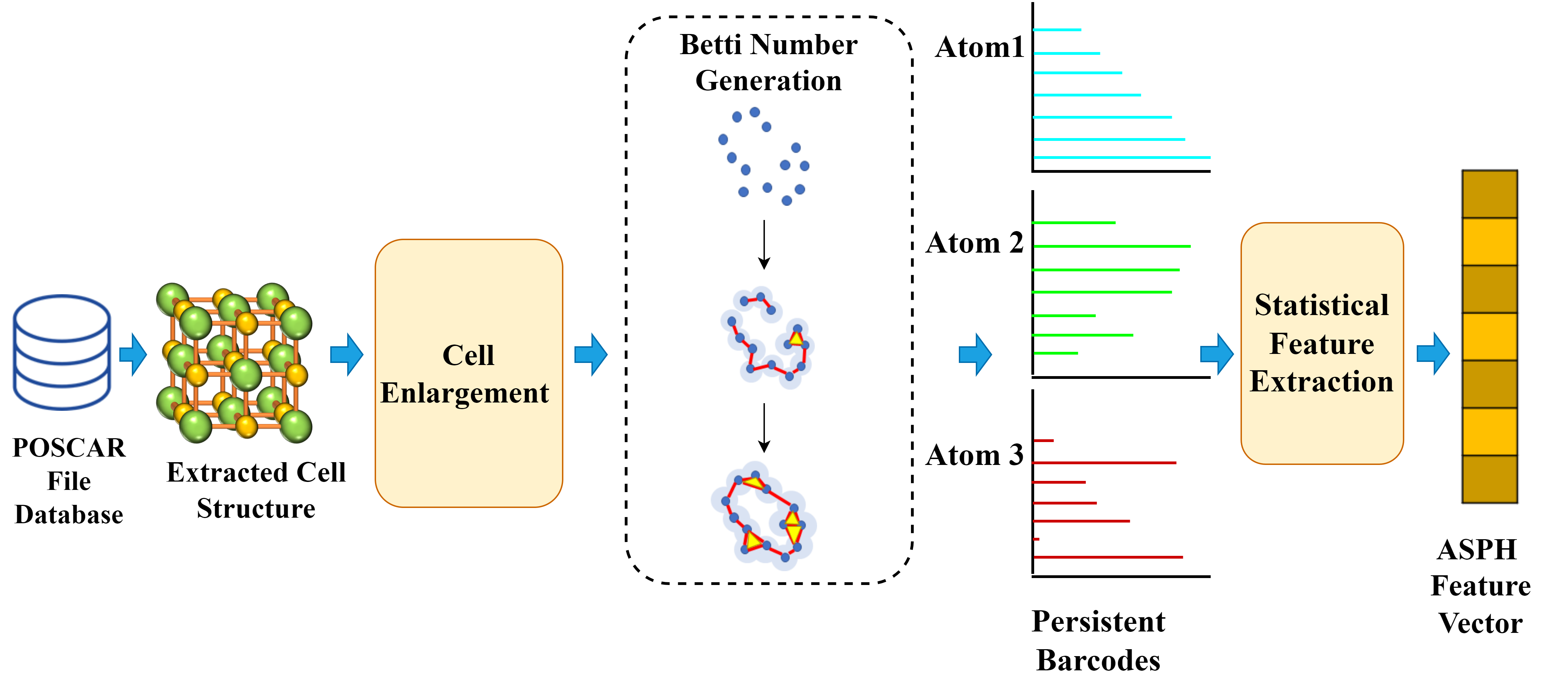}
\caption{Atom-Specific Persistent Homology (ASPH) pipeline implemented in parallel with the CGCNN pipeline. Crystal cell structure is extracted from the database followed by cell enlargement and Betti number generation. Later, persistent barcodes are produced to provide the topological ASPH feature vector.}
\label{ASPH pipeline}
\end{figure}

To better understand how ASPH combines atom-wise geometric and chemical information with the topological invariant to generate an individual topological fingerprint, first, we analyze the Betti number generation based on the simplicial homology group rank \cite{jiang2021topological}. To ease computational complexity, a group of simplex structures is used to describe a complex crystal shape instead of the original counterparts. Each simplicial complex is denoted by a k-simplex where k+1 is the number of affinely independent points, i.e., vertices in that simplex. From those simplices, the k-cycle group, $Z_k$, and k-boundary group ($B_k$) are formed. The resultant quotient group produced by $Z_k$ modulo $B_k$ is defined as the k-homology group, $H_k$. $H_k$ is more significant in this case because its rank denotes the desired $k^{th}$ Betti number; the persistent Betti numbers are directly used as the topological fingerprints in the later proceedings of the deep learning model's training.

Next, to extract element-specific interaction features, all the possible element pairs of the crystal composition are considered. Each element-specific pair collection of atoms inside a crystal is symbolized as $P^\beta_{\alpha,i}$, where $\alpha$ and $\beta$ are the two types of elements of that crystal composition, and the i$^{th}$ central atom of type-$\alpha$ is surrounded by atoms of type-$\beta$. Then, the unit cell is enlarged to the extent that the distance between any atom in the original unit cell and the boundary atom is smaller than a cutoff radius $r_{co}$. In this enlarged unit cell, a point cloud of atoms within $r_{co}$ is selected as the region of interest to generate the homology groups and persistence barcodes based on the Betti numbers. The point cloud region $R_i^{\alpha,\beta}$ is defined as \cite{jiang2021topological}:

\begin{equation}
R_{i}^{\alpha, \beta} = { r_{j}^{\beta} \: \| r_{i}^{\alpha}-r_{j}^{\beta} \| < r_{c}, r_{j}^{\beta},r_{i}^{\alpha} \in P_{\alpha, i}^{\beta} \forall j \in 1,2,.....N}.
\end{equation}

In our atom-specific persistent homology pipeline shown in Fig. \ref{ASPH pipeline}, crystal structure information from the corresponding Vienna Ab Initio Simulation Package (VASP) \cite{kresse1993ab,kresse1996efficiency} POSCAR files are gathered and their unit cells are enlarged to encompass a cutoff radius of 8{\AA}. Next, Betti numbers and persistent barcodes are generated for each of the structures using the Ripser package \cite{ctralie2018ripser}. Betti 0, 1, and 2 numbers and the persistent barcodes encode the topological feature of each atom in the structure. The topological features of an atom are conjoined with the neighboring atoms' topological features using five statistical quantities: minimum, maximum, mean, standard deviation, and the sum of each of the birth, death, and persistent length to produce atom-specific persistent homological feature vectors. For our case, these vectors are found to be 3115 dimensional, and most importantly, they are translation and rotation invariant.
\begin{figure}[h!]

\centering
\begin{subfigure}[b]{0.45\textwidth}
\centering
\includegraphics[width=\textwidth,height=7.8cm]{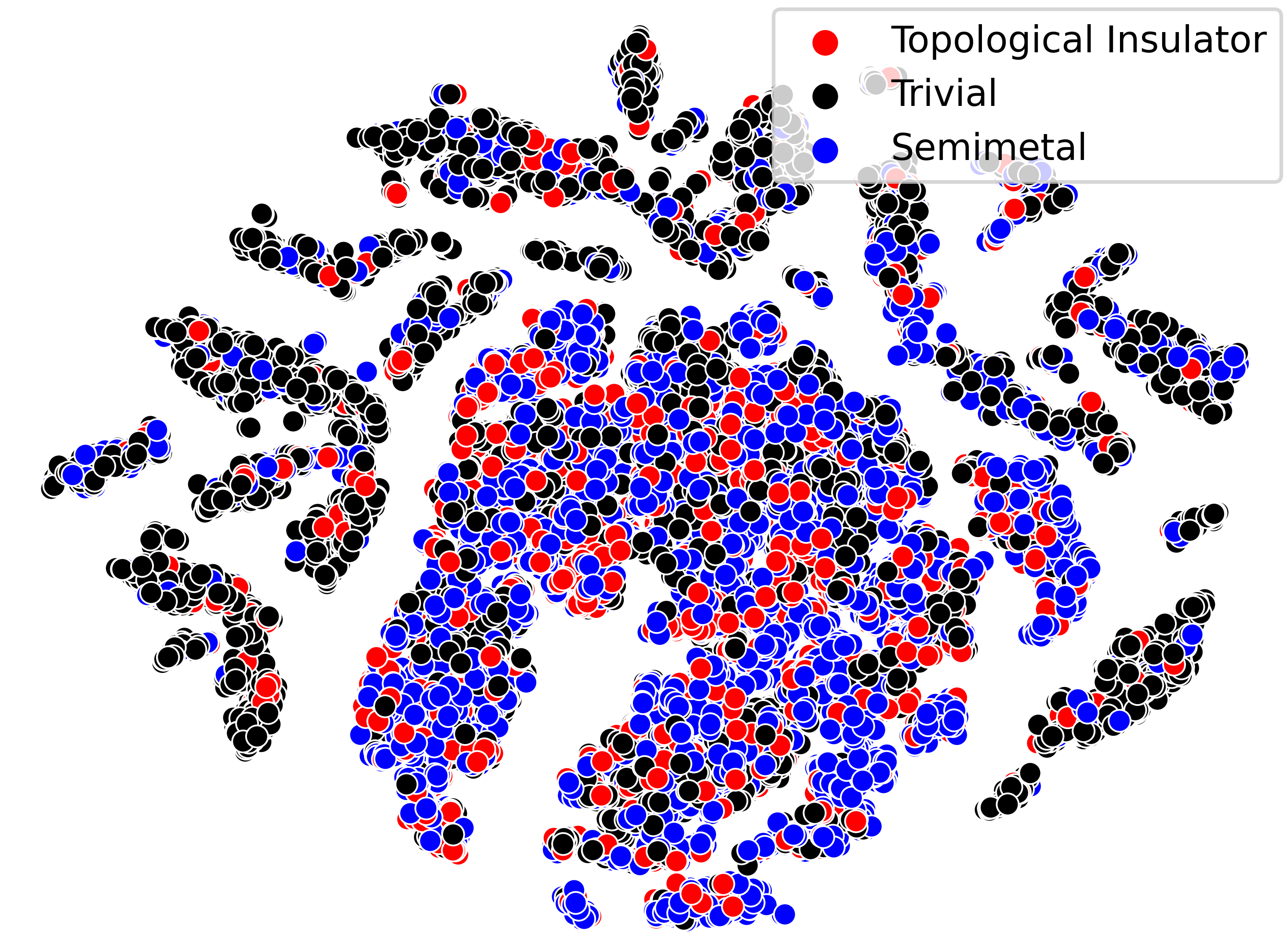}
\caption{}
\end{subfigure}
\hfill
\begin{subfigure}[b]{0.45\textwidth}
    \centering \includegraphics[width=\textwidth,height=7.8cm]{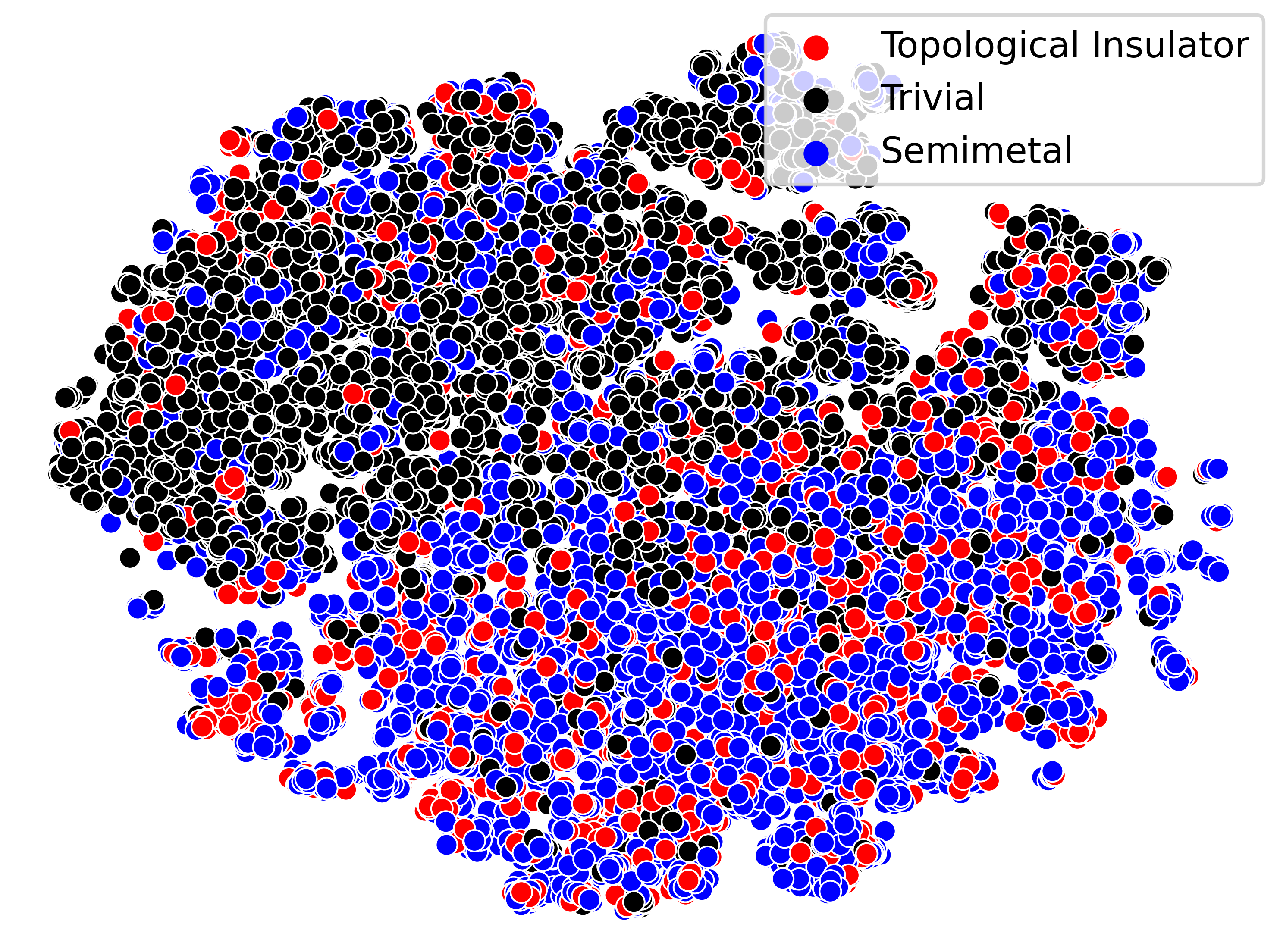}
\caption{}
\end{subfigure}
\hfill
\vspace{10pt}
\caption{Atom Specific Persistent Homology feature vectors projected in the two-dimensional plane using t-distributed stochastic neighbor embedding. In the figure, topological features are represented in the right (b) and chemical feature embeddings are displayed in the left (a).}
\label{ASPH feature vector}
\end{figure}

Figure \ref{ASPH feature vector} captures the topological high-dimensional feature vectors projected onto a two-dimensional plane with the help of the t-distributed Stochastic Neighbor Embedding (tSNE) algorithm. Intriguingly, topological semimetals, topological insulators, and trivial materials are segregated in the projection space. However, it is evident from the two figures that persistent homological features embody the differential distribution of data points that can aid in predicting the precise property of the material. It is also worth mentioning that, topological insulator data points are disseminated across the entire space. This observation justifies our point of utilizing topological descriptors for classifying materials into their respective topological or trivial class.

\vspace{5pt}
\begin{figure}[h!]
\centering\includegraphics[width=14cm]{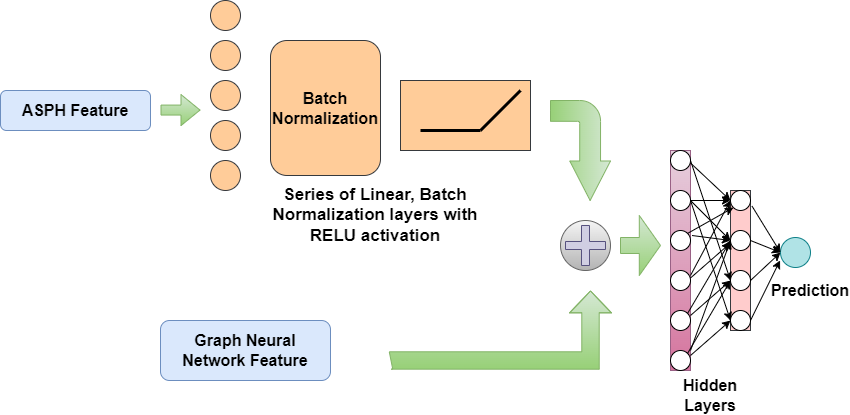}
\vspace{10pt}
\caption{Detailed implementation of our proposed model. The high dimensional persistent homological feature vector is reduced in dimension through a series of linear layers, followed by batch normalization and rectified linear activation layer. This vector is concatenated with a graph neural network vector and propagated through a deep neural network to produce the final result.}
\label{GCN-ASPH Composite Network}
\end{figure}

\subsection{Classification results}
The structural features obtained from the graph convolutional network and topological descriptors from atom-specific persistent homology are integrated and processed through a deep neural network to perform the final prediction. A detailed illustration of the ensemble model is depicted in Figure \ref{GCN-ASPH Composite Network}. We have tested our proposed model in predicting the three topological classes (trivial, semimetal, or topological insulator) and binary classes (topological vs. non-topological). The proposed model is trained on 80\% of the dataset and then it is applied on the test set, i.e., 20\% of the dataset comprising $\sim 2000$ materials. A detailed description of the methodology followed in this study is presented in Section \ref{sec:Methods}. The test set was completely unseen to our model during the training phase. For the test set, our proposed model yields an accuracy of 91.4\% on binary classification tasks and around 80\% on three-class classification with F1-scores around 88.5\% and 78.2\% respectively. This result outperforms all the state-of-the-art prediction models by around 7\% in accuracy (binary classification) and around 2\% in three-class classification tasks. The reason behind the disparate performance gains in binary and multi-class classification can be attributed to the persistent homological features of the proposed model. From Figure \ref{ASPH feature vector}(b), it is clear that persistent homological features can differentiate topological (semimetal and topological insulator) from non-topological(trivial) materials but it fails to segregate the three individual classes, which is evident from the topological insulators  (represented by red dots) spread all over the feature space. Examples of some materials from the two classes and the respective confidence scores are presented in Table \ref{tab:known_materials}. As seen in Table \ref{tab:known_materials}, the developed model can predict the true class of the material in question with a reasonably high confidence score. We also examine a few newly discovered materials, which were not included in the dataset [Table \ref{tab:recent_materials}] to test the ability of our model. It succeeds in predicting most of the materials with high confidence.

\begin{table}[h!]
\center
\caption{Classification performance on well-known materials.}
\begin{tabular}{c c c}
\hline
\textbf{Trivial material} & \textbf{\begin{tabular}[c]{@{}c@{}}Confidence\\ score\end{tabular}} \\ \hline
$\mathrm{CdWO_{4}}$  & 0.9949 \\ 
$\mathrm{GaTe}$   & 0.9976 \\ 
$\mathrm{YVO_{4}}$   & 0.9498 \\ 
$\mathrm{SF_{6}}$    & 0.9843 \\
$\mathrm{As_{4}S_{5}}$    & 0.9945 \\ 
$\mathrm{TePbF_{6}}$    & 0.9974 \\ 
$\mathrm{NaSbF_{4}}$    & 0.9962 \\ 
$\mathrm{HgTe}$    & 0.9710 \\ 
$\mathrm{Cs_{2}Cu_{2}Sb_{2}Se_{5}}$    & 0.9683 \\ 
$\mathrm{BaF_{2}}$   & 0.9970  \\ 
$\mathrm{LiEu_{3}O_{4}}$   & 0.9544  \\
$\mathrm{BaMgF_{4}}$   & 0.9999  \\
$\mathrm{CsTaI_{6}}$   & 0.9985  \\
$\mathrm{BaLaCaTe_{3}}$   & 0.9834  \\
$\mathrm{BaTiO_{4}}$   & 0.9999  \\
\hline
\end{tabular}
\quad
\begin{tabular}{c c c}
\hline
\textbf{Topological material} & \textbf{\begin{tabular}[c]{@{}c@{}}Confidence\\ score\end{tabular}} \\ \hline
$\mathrm{Sb_{2}Te_{3}}$ & 0.8650  \\ 
$\mathrm{LaSi}$   & 0.9163 \\ 
$\mathrm{Ca_{3}Ni_{7}B_{2}}$ & 0.9603 \\ 
$\mathrm{Sb_{2}Te_{3}}$ & 0.8650  \\
$\mathrm{LaH_{2}}$ & 0.9564  \\ 
$\mathrm{Ba_{5}Al_{5}Sn}$ & 0.9851  \\ 
$\mathrm{Pr_{3}Ga}$ & 0.9990  \\ 
$\mathrm{Bi_{2}Se_{3}}$& 0.6644 \\
$\mathrm{DyNbO_{4}}$ & 0.9890  \\ 
$\mathrm{LuNi_{2}Sn}$ & 0.9375  \\
$\mathrm{FeB}$ & 0.9098  \\
$\mathrm{Nb_{3}Ga}$ & 0.992  \\
$\mathrm{LuTiSi}$ & 0.9915  \\
$\mathrm{LiVS_{2}}$ & 0.8794  \\
$\mathrm{Dy_{2}Fe_{2}Si_{2}C}$ & 0.9705  \\
\hline 
\end{tabular}
\label{tab:known_materials}
\end{table}

\begin{table}[h!]
\center
\caption{Classification performance on newly discovered materials.}

\begin{tabular}{c c c}
\hline
\textbf{Trivial material} & \textbf{\begin{tabular}[c]{@{}c@{}}Confidence\\ score\end{tabular}} \\ \hline
$\mathrm{Bi_{4}I_{4}}$  &   0.9982
          \\ 
 $\mathrm{Sr_{2}SnO_{4}}$  &   0.9989
          \\ 
 $\mathrm{RbLa(MoO_{4})_{2}}$  &   0.8991
          \\ 
   $\mathrm{P_{4}PtF_{12}}$  &   0.9129
          \\
$\mathrm{GaAg(PSe_{3})_{2}}$  &   0.9990 \\
$\mathrm{K_{3}ReC_{4}(N_{2}O)_{2}}$  &   0.999 \\
$\mathrm{HgHNO_{4}}$   & 0.9999  \\
$\mathrm{LiCaAs}$  &   0.9999 \\
$\mathrm{KPt_{2}Se_{3}}$  &   0.9831 \\
$\mathrm{LiCuO}$  &   0.9225 \\
$\mathrm{PAuS_{4}}$  &   0.8890 \\
$\mathrm{LuH_{2}ClO_{2}}$  &   0.9970 \\
          \hline
\end{tabular}
\quad
\begin{tabular}{c c c}
\hline
\textbf{Topological material} & \textbf{\begin{tabular}[c]{@{}c@{}}Confidence\\ score\end{tabular}} \\ \hline
$\mathrm{RbV_{3}Sb_{5}}$      &  0.7433
          \\ 
$\mathrm{In_{2}Ni_{3}Se_{2}}$       &    0.999        \\ 
$\mathrm{CsV_{3}Sb_{5}}$      &  0.7341
          \\ 
$\mathrm{KV_{3}Sb_{5}}$    &  0.7441
          \\ 
$\mathrm{ZrPRu}$    &  0.9473
          \\ 
$\mathrm{Pr(GeRu)_{2}}$    &  0.9946
          \\ 
$\mathrm{CeAu}$    &  0.9820
          \\
$\mathrm{VAsRh}$    &  0.9985
          \\ 
$\mathrm{Ho(CuGe)_{2}}$    &  0.9943
          \\ 
$\mathrm{Ti_{2}BRh_{6}}$      &  0.9999
          \\
$\mathrm{EuAu_{5}}$    &  0.9999
          \\ 
$\mathrm{Sr_{2}TbReO_{6}}$    &  0.9049
          \\ 
          \hline
\end{tabular}
\label{tab:recent_materials}
\end{table}

\begin{figure}[t!]
\centering
\begin{subfigure}{0.85\textwidth}
\includegraphics[width=\textwidth,height=7.8cm]{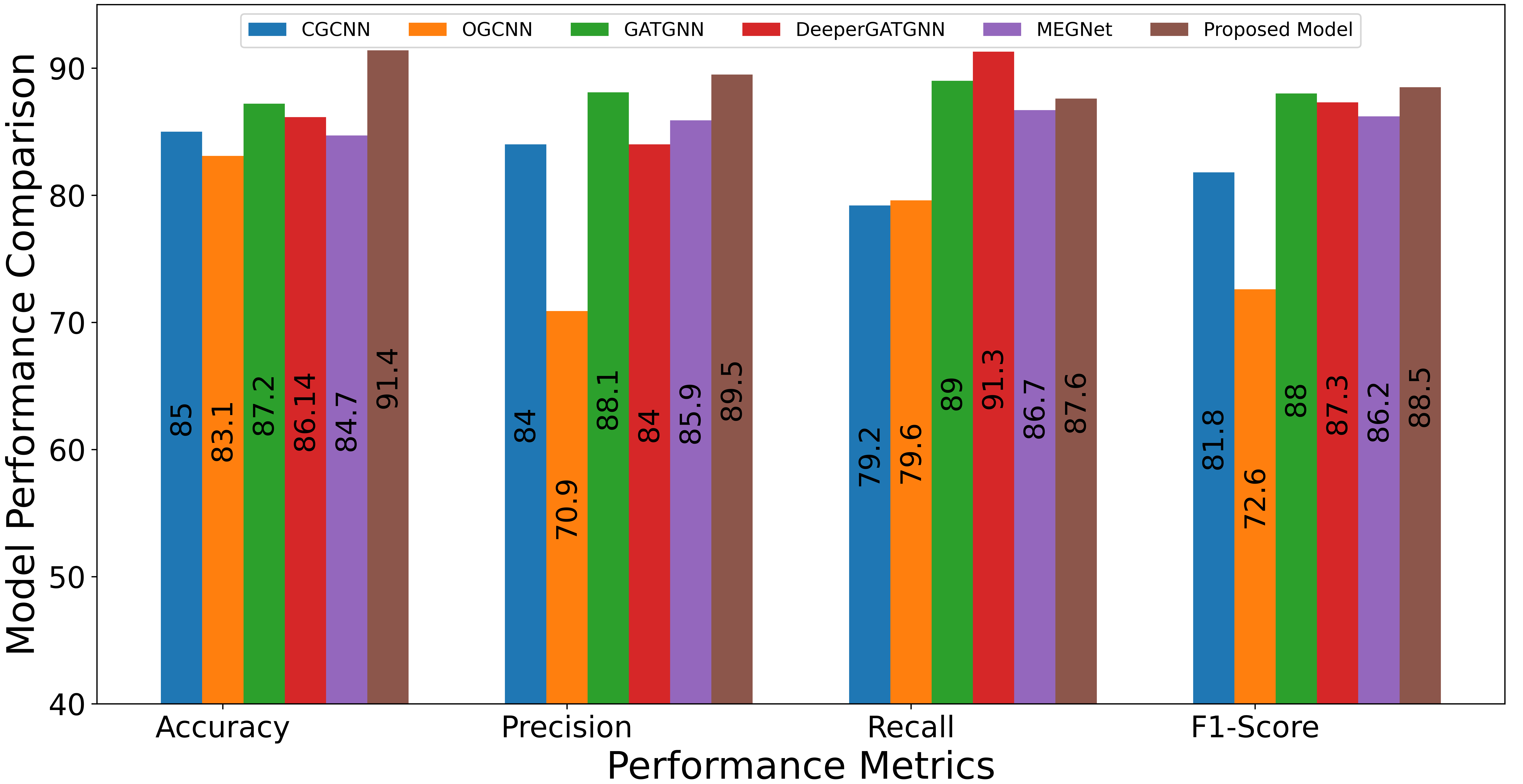}
\caption{}
\end{subfigure}

\begin{subfigure}{0.85\textwidth}
\includegraphics[width=\textwidth,height=7.8cm]{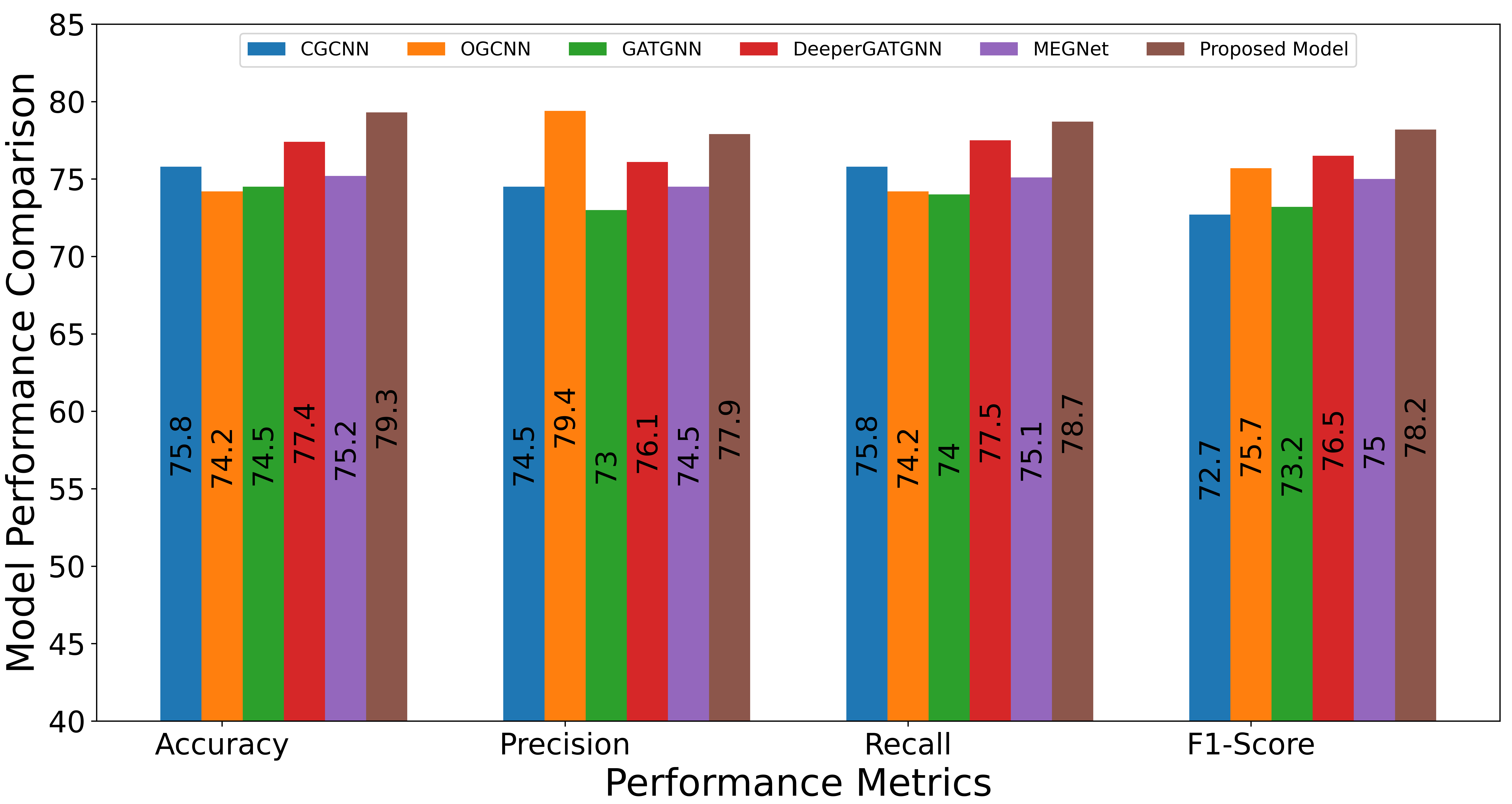}
\caption{}
\end{subfigure}
\caption{Comparison of model performance in predicting topological class in different classification metrics. (a) shows the performance of different models in binary classification (topological vs trivial) while (b) shows the performance in three-class classification task (trivial, topological semimetal, and topological insulator)}
\label{Bar}
\end{figure}

In order to compare the proposed model with some other state-of-the-art classifiers, we calculate accuracy, precision, recall, and F1 scores on the test set for our developed model as well as the crystal graph convolutional neural network (CGCNN), orbital graph convolutional neural Network (OGCNN), GATGNN\cite{gatgnn}, DeeperGATGNN\cite{deepergatgnn} and Materials Graph Network (MEGNet)\cite{megnet}. Fig. \ref{Bar} captures the quantitative comparisons between these models. Applying OGCNN and CGCNN on our test set yields an accuracy of 83\% and $85\%$, respectively. Enabled by the combination of atom-specific information and CGCNN, our model scores a $91.4\%$ accuracy on the testing set, outperforming the CGCNN and OGCNN by 6.4\% and 8.3\%, respectively. Moreover, transformer and attention-based models have been the center of attention for quite some time. Our proposed model is seen to surpass graph attention models such as GATGNN\cite{gatgnn} and DeeperGATGNN\cite{deepergatgnn} by an appreciable margin. The precision, recall, and F1 scores are also improved significantly in our proposed model compared to the other models. We expect our model to have twofold impacts: first, it will enable an efficient classification and search for new topological materials. Second, it will encourage the development of new machine-learning-based classifiers/discriminators incorporating atom-specific features to enable faster material search for physics, chemistry, and biological applications.

\section{Discussion}
In summary, our proposed model contributes to the exploration of novel topologically reactive materials by exploiting persistent homological features and integrating them with the graph convolutional neural network. The novelty of this work encompasses robust prediction of the topological nature of crystals along with efficient incorporation of graph features with persistent homological information. In spite of the fact that graph neural networks have been rigorously used in material property prediction, the impact of graph representation and graph convolutional features in encoding topological information of material has been an untrodden realm. It is observed in this work that, structural information and neighborhood connectivity information encoded in the graphs of the materials can be employed to predict the topological nature of materials. Furthermore, manipulating the graph information with atom-specific persistent homological features of individual materials can stimulate the prediction. Crystals, being unstructured non-euclidean data, are difficult to assimilate in the deep learning pipeline. Graph feature along with persistent homological feature is an effective method to do this, reflecting upon the fact that these features are rotation and translation invariant. The proposed model has been tested against state-of-the-art graph neural network prediction models and showed promising results with an accuracy of over 91\% and an F1 score of over 89\%. Additionally, the model's performance has been examined on both notable and newly discovered topological materials, thus bearing testimony to the robustness of our model. Considering recent advancements in topological material physics and its anticipated application in the field of spintronics, optics, and electronics, our model presents a novel approach to studying topological materials bypassing time and resource-intensive density functional theory calculations and is expected to be a pioneer in the search for new topologically exquisite materials.

\section{Methods}
\label{sec:Methods}
\subsection{Dataset}
\begin{figure}[h!]
\centering\includegraphics[width=0.95\textwidth, height=12cm]{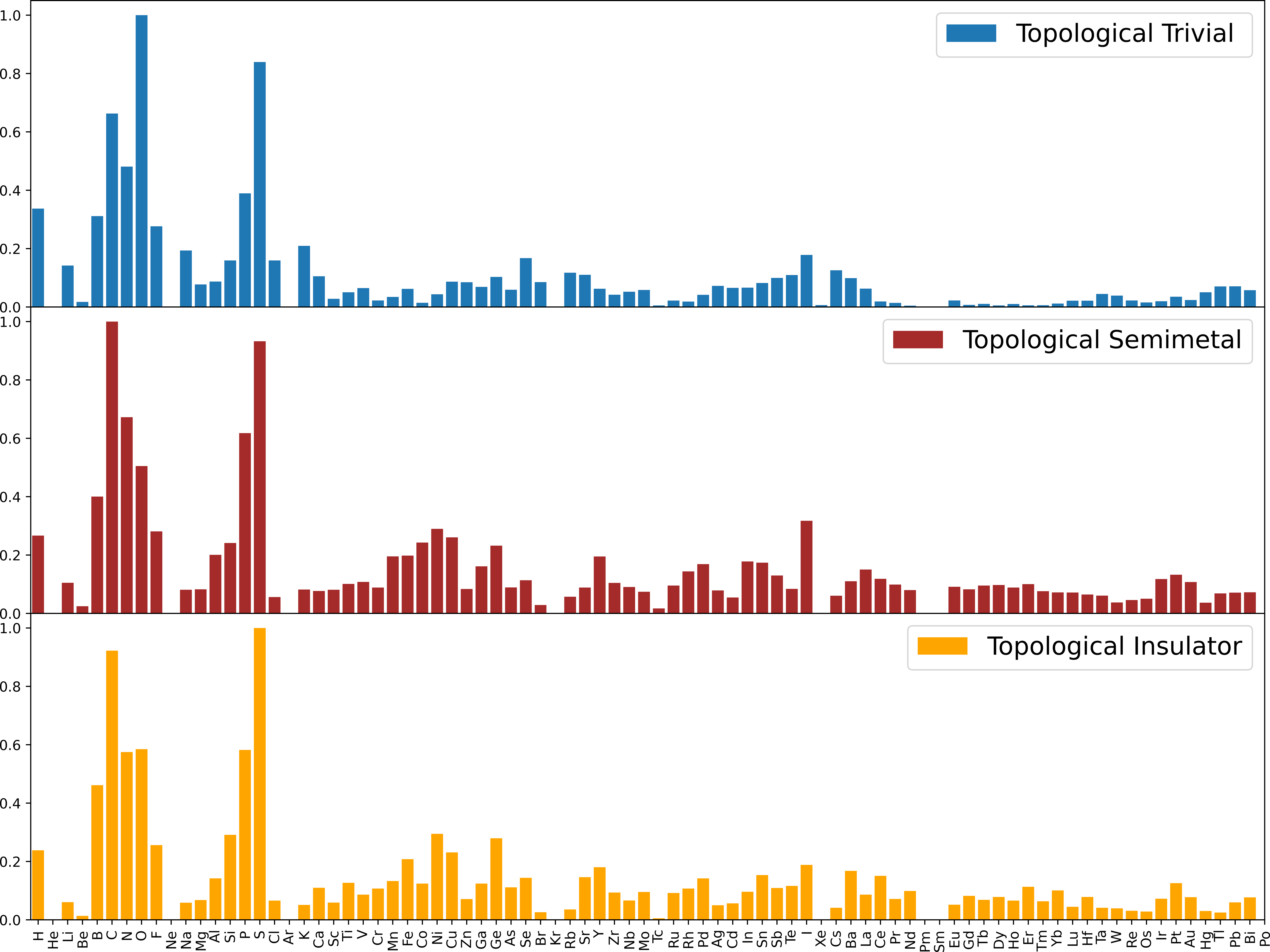}
\caption{Histogram plot demonstrating the elemental distribution in topological semimetals, trivials, and topological insulators.}
\label{barplot_dataset}
\end{figure}

\begin{figure}[h!]
\centering
\begin{subfigure}[b]{0.45\textwidth}
\centering
\includegraphics[width=0.9\textwidth,height=7.8cm]{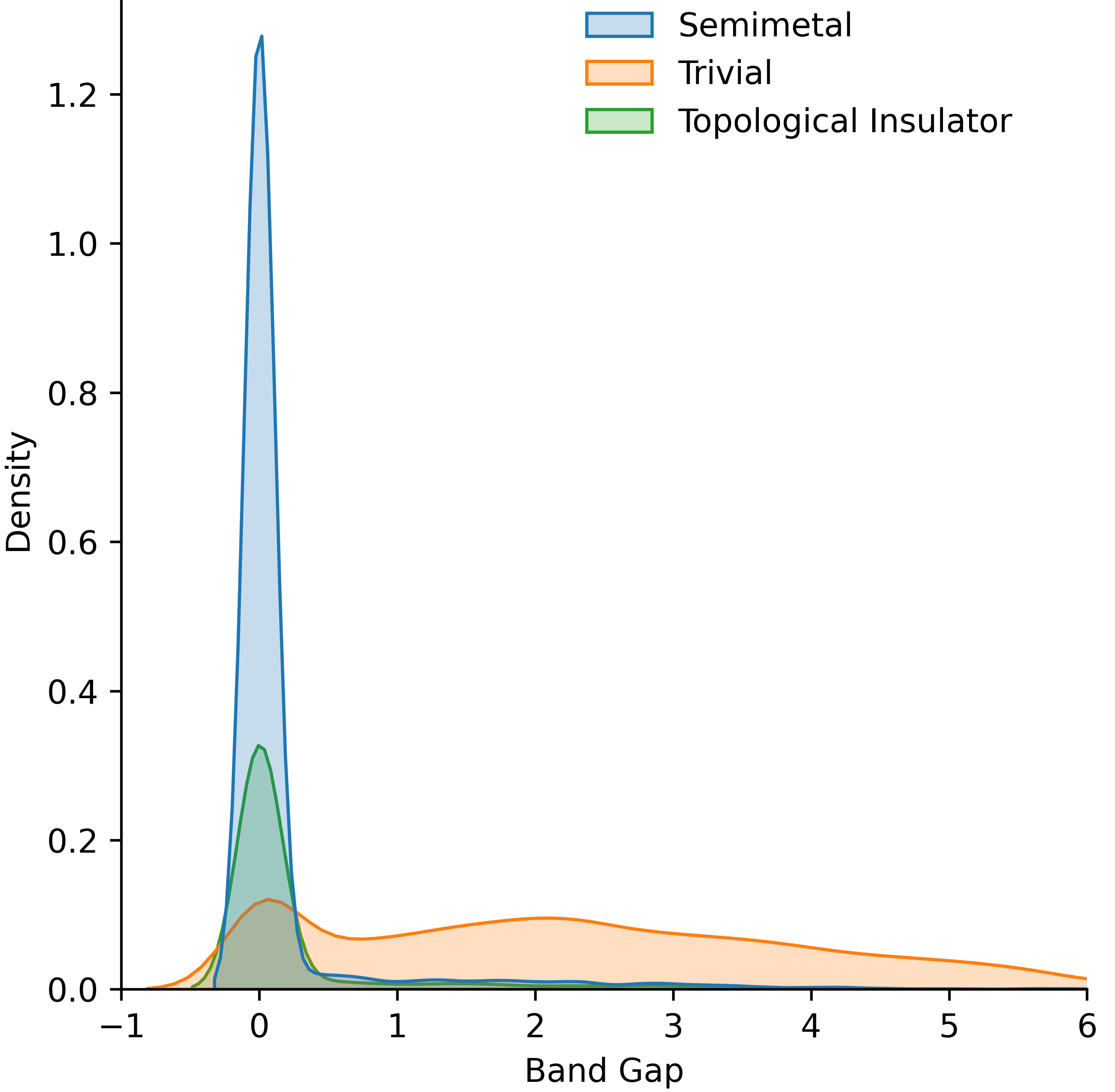}
\caption{}
\end{subfigure}
\hfill
\begin{subfigure}[b]{0.45\textwidth}
\centering 
\includegraphics[width=\textwidth,height=7.8cm]{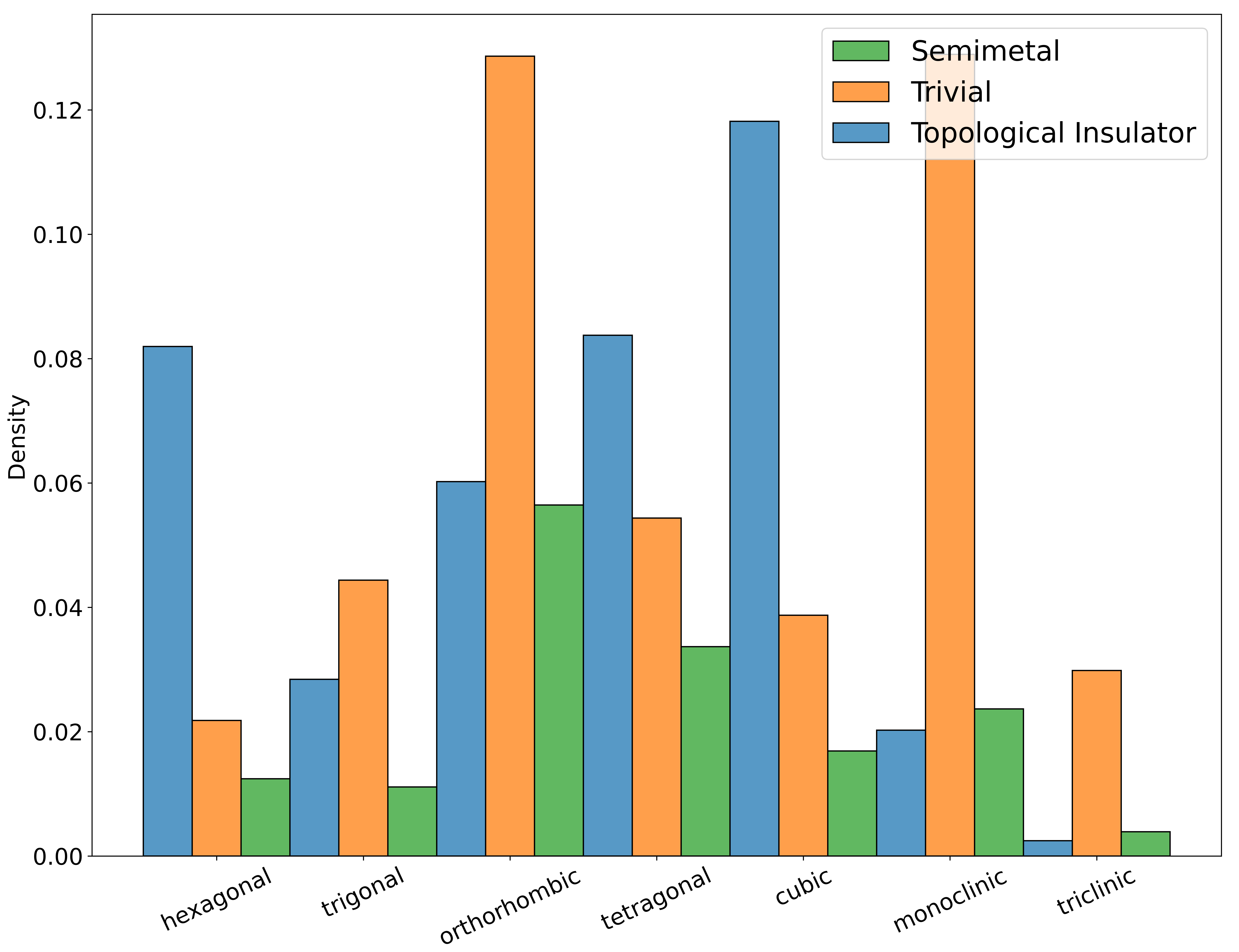}
\caption{}
\end{subfigure}
\hfill
\vspace{5pt}
\caption{Statistical analysis performed on topological and trivial materials. The distribution of band gap is exhibited in (a), while (b) depicts the variation in crystal structure for the three material classes.}
\label{stats}
\end{figure}

The dataset for this study has been collected from the Topological Quantum Chemistry database \footnote{https://www.topologicalquantumchemistry.org/, https://www.cryst.ehu.es/} \cite{bradlyn2017topological,vergniory2019complete,vergniory2021topological}. The final dataset consists of 33800 material entries: around 15000 topologically trivial materials, 5600 topological insulators, and 13250 topological semimetals. The topological material database is collected through web scrapping from the topological quantum chemistry website. Crystallographic Information File (CIF) files \cite{hall_crystallographic_1991} along with other physical and topological properties such as space group, topological invariant, etc. were also collected for input. The collected dataset was cleansed to eliminate the empty CIF files and erroneous parameters. Next, the dataset was converted to Atomic Simulation Environment format VASP POSCAR for compatibility with our model. The class distribution of topological trivial materials, semimetals, and insulators is demonstrated in Figure \ref{barplot_dataset}. It can be observed from the histogram plot that topological and trivial materials are statistically distinct in terms of constituent atoms that topological semimetals and topological insulators have a high probability of containing d-block elements. 
\vspace{5pt}
Further insight can be obtained from Figure \ref{stats} regarding the physically differentiable characteristics of topological and trivial materials. Figure \ref{stats}(a) and \ref{stats}(b) depict the discernible attributes in the bandgap and crystal structures between trivial and non-trivial materials. Evidently, topological materials are low-bandgap materials with predominant cubic or hexagonal crystal symmetry. On the other hand, topologically trivial materials show a probability of having an orthorhombic or monoclinic structure with a wide range of bandgaps. 

\subsection{Implementation Details}
The graph convolutional neural network and atom-specific persistent homology-based ensemble model is implemented in PyTorch \cite{NEURIPS2019_9015}. Pytorch Geometric \cite{fey2019fast}, a specific library to design and implement graph neural networks has been adopted in this work. For the graph convolutional neural network, atom feature-length and neighbor feature length have been taken as 64 with 1 hidden layer and 128-dimensional hidden feature. Maximum 12 neighbors with a cutoff radius of 15 {\AA} are considered in the generation of graphs. 3 graph convolutional layers are incorporated in the model and they are designed with concatenated atomic feature and neighbor atom feature being passed through a fully connected layer, batch normalization layer with succeeding gating function, and sigmoid activation function.

In the persistent homological feature generation process, given cells are enlarged to a cutoff radius of 8{\AA} and the betti number for each structure is calculated accordingly. Persistent barcodes of the dataset are generated using Ripser package \cite{ctralie2018ripser} and five statistical features i.e. minima, maxima, average, standard deviation, and the summation of birth, death, and persistent length are extracted. In the case of our implementation, a 3115-dimensional persistent homological feature is produced for each material and these features are processed through a series of fully connected layers, batch normalization layers, and rectified linear unit activation layers, eventually reducing to a dimension of 64 and merged with output features of graph convolutional neural network. Dropout layers are used as regularizers to ensure homogeneity of information distribution among the units. Concatenated features of dimension 192 are finally passed through a shallow classification layer with a log softmax activation function to determine the topological class of each material.

The model has been instructed to minimize the Negative Log-Likelihood loss function \cite{THEODORIDIS2020301} using the Adaptive Moment Optimization (Adam) optimizer \cite{kingma2014adam} with a learning rate of 0.001 and zero weight decay and trained for 50 epochs. The learning rate is controlled to reach the minima in the loss landscape using step learning rate scheduling with 10 10-fold reduction in learning rate in reaching 30 and 40 epoch milestones. Accuracy, precision, recall, and F1 scores for model evaluation have been calculated using the Scikit-learn library \cite{scikit-learn}.

\subsection{Training Scheme}
The deep learning models of this work have been implemented on computers of the Robert Noyce Simulation Laboratory at Bangladesh University of Engineering and Technology. The proposed model along with the reference models are trained on 8 core Intel Core i7 CPU and Nvidia GTX 1650 for 50 epochs with batch size 256. The dataset is split into 80-10-10 ratios for training, testing, and validation purposes, respectively, and therefore experienced a total of 26000 training samples. Dataset validation has also been applied to ensure stability and robustness of results, whereas dropout layers are introduced in order to avoid overfitting. \cite{srivastava2014dropout}.  

\section{Acknowledgements}
Financial grant to procure VASP is given by the Bangladesh University of Engineering and Technology (BUET). Computational facilities provided by BUET and Princeton University are duly acknowledged. Work at Princeton University is supported by the Gordon and Betty Moore Foundation (GMBF4547 and GMBF9461).

\section{Author Contributions}
M.S.H. and Q.D.M.K. conceived the project. A.R. and A.G.D. initiated the research work presented here. A.R. designed and trained the models. The analysis has been performed by A.R., A.G.D., and M.S.H. A.G.D., H.R., and A.R. prepared the draft with contributions from M.S.H., M.Z.H., and Q.D.M.K. The project has been supervised by M.S.H., M.Z.H., and Q.D.M.K.

\section{Conflict of Interest}
The Authors declare no Competing Financial or Non-Financial Interests.

\section{Data Availability}
The data that support the findings of this study are available on the website \url{https://topologicalquantumchemistry.com/}. Furthermore, the complete dataset can be obtained from the corresponding author upon reasonable request.

\section{Code Availability}
The code, model description, and training configurations used in this work will be available on GitHub.

\bibliography{main}

\end{document}